\documentclass[12pt,preprint]{aastex}

\slugcomment{Printed: \today}

\shorttitle{HCO$^+$ in K~3-35}
\shortauthors{Tafoya et al.}

\begin{document}


\title{Detection of HCO$^+$ Emission toward the Planetary Nebula K~3-35}


\author{D. Tafoya\altaffilmark{1,2}, Y. G\'omez\altaffilmark{1}, G.
Anglada\altaffilmark{3}, L. Loinard\altaffilmark{1}, J. M.
Torrelles\altaffilmark{4}, L. F. Miranda\altaffilmark{3}, M.
Osorio\altaffilmark{3}, R. Franco-Hern\'andez\altaffilmark{1,2}, L-\AA,
Nyman\altaffilmark{5}, J. Nakashima\altaffilmark{6} and S.
Deguchi\altaffilmark{7}}






\altaffiltext{1}{Centro de Radioastronom\'\i a y Astrof\'\i sica, UNAM,
Apartado Postal 3-72 (Xangari), 58089
Morelia, Michoac\'an, M\'exico; d.tafoya@astrosmo.unam.mx, y.gomez@astrosmo.unam.mx, l.loinard@astrosmo.unam.mx, r.franco@astrosmo.unam.mx}
\altaffiltext{2}{Harvard-Smithsonian Center for Astrophysics, 60 Garden 
Street, Cambridge, MA 02138; dtafoya@cfa.harvard.edu, rfranco@cfa.harvard.edu}
\altaffiltext{3}{Instituto de Astrof\'\i sica de Andaluc\'\i a, CSIC,
Apartado 3004, E-18080, Granada, Spain; guillem@iaa.es, lfm@iaa.es, osorio@iaa.es}
\altaffiltext{4}{Instituto de Ciencias del Espacio (CSIC) and Institut
d'Estudis Espacials de Catalunya, Facultat de F\'{\i}sica, Planta 7a,
Universitat de Barcelona, Av. Diagonal 647, 08028 Barcelona, Spain; torrelles@ieec.fcr.es}
 \altaffiltext{5}{Swedish-ESO Submillimetre Telescope, European Southern
Observatory, Casilla 19001, Santiago 19, Chile; Onsala Space Observatory,
439 92 Onsala Sweden; lnyman@eso.org}
 \altaffiltext{6}{Institute of Astronomy and Astrophysics, Academia 
Sinica, Taipei, Taiwan; junichi@asiaa.sinica.edu.tw}
 \altaffiltext{7}{Department of Astronomical Science, Graduate University
for Advanced Studies, Nobeyama Observatory, Minamimaki, Minamisaku, Nagano
384-1305; deguchi@nro.nao.ac.jp}


\begin{abstract}
 We report the detection, for the first time, of HCO$^+$ 
($J=1\rightarrow0$) emission as well as marginal CO ($J=1\rightarrow0$) 
emission toward the planetary nebula (PN) K~3-35 as a result of a 
molecular survey carried out toward this source. We also report new 
observations of the previously detected CO ($J=2\rightarrow1$) and water 
maser emission, as well as upper limits for the emission of the SiO, 
H$^{13}$CO$^{+}$, HNC, HCN, HC$_{3}$OH, HC$_{5}$N, CS, HC$_{3}$N, 
$^{13}$CO, CN, and NH$_{3}$ molecules. From the ratio of CO 
($J=2\rightarrow1$) to CO ($J=1\rightarrow0$) emission we have estimated 
the kinetic temperature of the molecular gas, obtaining a value of 
$\simeq$ 20 K. Using this result, we have estimated a molecular mass for 
the envelope of $\simeq$ 0.017 $M_{\odot}$, and an HCO$^{+}$ abundance 
relative to H$_2$ of $6\times 10^{-7}$, similar to the abundances found in 
other PNe. K~3-35 is remarkable because it is one of the two PNe reported 
to exhibit water maser emission, which is present in the central region as 
well as at a distance of $\simeq$ 5000 AU away from the center. The 
presence of molecular emission provides some clues that could help to 
understand the persistence of water molecules in the envelope of K 3-35. 
The HCO$^{+}$ emission could be arising in dense molecular clumps, that 
may provide the shielding mechanism which protects water molecules in this 
source.

\end{abstract}


\section{Introduction}

The chemical composition of the molecular envelope that surrounds a young 
planetary nebula (PN) can reflect the recent history of the transition from 
the asymptotic giant branch (AGB) to the PN phase. When envelopes of AGB stars 
are oxygen-rich, they may produce strong maser emission of one or more molecules 
such as OH, H$_{2}$O, or SiO, which commonly appears stratified, with the SiO masers 
located close to the stellar surface, water masers in the range of about 10-100 AU, 
and OH masers farther away, up to 10$^{4}$ AU from the central star \citep{R81,CC86}. 

Molecules that were present in the red giant envelope are destroyed by the 
radiation of the core as the star evolves to the PN phase, while other 
molecular species could develop in these peculiar physical conditions. In 
particular, water-vapor masers that are detected in the giant envelopes 
\citep{R81,E92,H96}, and also in some proto-PNe 
\citep{LM88,MB99,GR02,im02}, are not expected to persist in the PN phase, 
where the envelope not only begins to be ionized but also becomes rarefied 
(Lewis 1989; G\'omez, Moran \& Rodr\'\i guez 1990). As the slow and 
massive mass loss of the late AGB-phase stops and the star enters in its 
PN phase, the water molecules are expected to disappear in a time scale of 
decades \citep{GMR90}, and only OH masers seem to persist for a 
considerable time ($\simeq$ 1000 yr; Kwok 1993). However, recently two PNe 
(K~3-35 and IRAS~17347-3139; Miranda et al. 2001, de Gregorio-Monsalvo et 
al. 2004) have been found to harbor water maser emission, suggesting that 
these objects are in an early stage of their evolution as PNe, where the 
physical and chemical conditions could still permit the existence of water 
molecules.

K~3-35 is a very young PN (which evolved from an oxygen rich AGB star) characterized 
by an S-shaped radio continuum emission morphology with a well defined point-symmetric 
structure (Aaquist \& Kwok 1989; Aaquist 1993; Miranda et al. 2000, 2001). The 
detected water-vapor masers are located at the center of the nebula, apparently 
tracing a torus-like structure with a radius of $\simeq$~85~AU (adopting a distance 
of $\simeq$ 5 kpc, Zhang 1995), and in addition they are also found at the 
surprisingly large distance of 5000~AU from the star, in the tips of the bipolar 
lobes \citep{M01,G03}. Water masers were not expected to be found at such an 
enormous distance in an evolved star since the physical conditions required to pump 
the maser emission ($n_{\rm H_2}\simeq10^8$ cm$^{-3}$, $T_{\rm k}\simeq 500$ K) are 
difficult to explain, given our current understanding of these objects. \cite{M01} 
proposed that shocks driven by the bipolar jet could create the physical conditions 
necessary to pump the distant water-vapor masers. However, the presence and 
persistence of water molecules in these PNe are still puzzling, probably being related 
to some shielding mechanism due to the presence of high density molecular gas that 
protects water molecules against the ionizing radiation of the central star. 

To investigate the possible existence of such a shielding mechanism, one 
must characterize the physical conditions of the molecular gas in the 
envelope surrounding K~3-35. Interestingly, weak broad 
CO ($J=2\rightarrow1$) emission has been detected toward K~3-35 by 
\cite{DB96}, and more recently by \cite{Hugg05}, indicating the presence 
of a neutral molecular envelope and suggesting the possible presence of 
dense molecular clumps. In this paper, we present a search for additional 
molecular emission toward K~3-35, reporting the first detection of HCO$^+$ 
in this PN, confirmation of CO ($J=2\rightarrow1$) and water maser 
emission, as well as a tentative detection of CO ($J=1\rightarrow0$).

\section{Observations}

The survey for molecular lines toward the young planetary nebula K~3-35 was 
carried out using the Onsala 20 m telescope in Sweden, the 45 m Nobeyama telescope 
in Japan and the IRAM 30 m telescope in Spain (see Table \ref{table_1}). The 
coordinates RA(J2000)=$19^h 27^m 44^s$, Dec(J2000)=$21^\circ 30' 03''$ were adopted
for the position of K~3-35 \citep{M01}. In what follows, we describe each of these 
observations. 

\subsection {20 m Onsala}

Observations of several molecular transitions from 86 to 113 GHz were 
carried out using the 20 m telescope of the Onsala Space Observatory 
(OSO)\footnote[8]{Onsala Space Observatory at Chalmers University of 
Technology is the Swedish National Facility for Radio Astronomy} in the 
dual beam switching mode. We used a 100 GHz receiver which includes a 
cryogenically cooled SIS-mixer covering the frequency range from 84 to 116 
GHz. The backend used was the LCOR autocorrelator which, for all 
molecular transitions except for $^{13}$CO, had a bandwidth of 80~MHz, 
providing a resolution of 50 kHz. This corresponds to a velocity resolution 
of $\simeq$ 0.17 km~s$^{-1}$ at 86 GHz and $\simeq$ 0.13 km~s$^{-1}$ at 115 GHz. 
The $^{13}$CO observations were carried out with a bandwidth of 40~MHz, 
providing a spectral resolution of 25 kHz, which corresponds to a velocity resolution 
of $\simeq$~0.068~km~s$^{-1}$. In all cases the bandwith was centered at 
$v_{\rm LSR}$~=~20~km~s$^{-1}$. The half power beam width (HPBW) of 
the telescope ranges from $\simeq$ 44$^{\prime \prime}$ at 86 GHz to 
33$^{\prime \prime}$ at 115~GHz. System temperatures were typically in the 
range from 300 to 500 K. The parameters of the molecular transitions 
observed as well as additional telescope parameters are given in 
Table~\ref{table_1}.

\subsection {45 m Nobeyama}

Observations with the 45 m radio telescope at Nobeyama\footnote[9]{The 
Nobeyama Radio Observatory is a branch of National Astronomical 
Observatory, an interuniversity research institute, operated by Ministry 
of Education, Science, and Culture, Japan} were carried out using the 
position switching mode. The BEam Array Receiver System (BEARS), 
consisting of a grid of 5$\times$5 beams separated by $\sim41''$, was used 
for the observations of the CO ($J=1\rightarrow0$) transition at 115~GHz. 
The backend bandwidth was 32~MHz, and was centered at $v_{\rm 
LSR}$~=~20~km~s$^{-1}$, providing a spectral resolution of 30~kHz 
($\simeq$ 0.08 km~s$^{-1}$). The HPBW of the telescope at this frequency 
is $\simeq$ 16$''$. Typical system temperatures for the 115~GHz 
observations were in the 420-450~K range.

Observations of the H$_2$O maser and NH$_3$(1,1) lines at $\nu$ $\simeq$ 22-24 GHz were carried out using the H22 receiver. We used a bandwidth of 40~MHz centered at $v_{\rm LSR}$ = 20~km~s$^{-1}$ with a frequency resolution of 20 kHz ($\simeq$~0.25 km~s$^{-1}$). The HPBW of the telescope at these frequencies is $\sim70''$. Typical system temperatures for these observations were 200-300~K. The parameters of the molecular transitions observed as well as additional telescope parameters are given in Table~\ref{table_1}.

\subsection {30 m IRAM}

Observations of the CO ($J=2\rightarrow1$) line at 230 GHz were carried
out with the IRAM 30 m telescope at Pico Veleta using the position
switching mode. We used the multibeam HEterodyne Receiver Array (HERA),
which consists of nine receivers arranged in a regular 3$\times$3 grid
with spacing on the sky of 24$''$, and a beam size of $\sim12''$. The
backend used was the VErsatile Spectral and Polarimetric Array (VESPA)
autocorrelator splitted into two parts, providing 438 channels of 0.41
km~s$^{-1}$ width, and 224 channels of 1.63 km~s$^{-1}$ width for each of
the nine receivers. The system temperature was $\sim$ 300~K, and the
pointing uncertainty was $\sim2''$. The parameters of the molecular
transition observed as well as additional telescope parameters are given
in Table \ref{table_1}.

\section{Results}

In Table \ref{table_1} we summarize the results of our molecular survey 
toward K~3-35. We succeeded in detecting weak emission of HCO$^{+}$ 
($J=1\rightarrow1$) and CO ($J=1\rightarrow0$), both reported for the 
first time in this work. We also detected CO ($J=2\rightarrow1$) emission, 
which was reported previously by \cite{DB96} and \cite{Hugg05}, and water 
maser emission, reported previously by Engels et al. (1985), Miranda et 
al. (2001), and de Gregorio-Monsalvo et al. (2004). The remaining 
molecular transitions listed in Table~\ref{table_1} have not been 
detected. The integrated intensity, central velocity, and line width of 
the detected lines are presented in Table \ref{table_1}, where we also 
give the rms achieved in all the transitions observed in our survey.

\subsection{H$_{2}$O}

Using the 45 m Nobeyama telescope we detected the H$_2$O
($J=6_{16}\rightarrow5_{23}$) maser line toward K~3-35 (see Fig. 1 and
Table~\ref{table_1}). The maser emission appears in the velocity range from 20 to
24~km~s$^{-1}$, with two clear components centered at $v_{\rm LSR}$~=~21.4
and 22.5~km~s$^{-1}$, similar to the velocities of the strongest peaks
observed by Engels et al.  (1985) and Miranda et al. (2001). The Very
Large Array (VLA) observations of Miranda et al. (2001) reveal that this
velocity range corresponds essentially to the features observed toward the 
center of K~3-35 and the tip of its northern lobe.
 
\subsection{CO}

Using the IRAM 30 m telescope we have detected CO ($J=2\rightarrow1$)
emission toward K~3-35. The observed emission has two velocity components,
a narrow component ($\Delta v~\simeq$~5~km~s$^{-1}$), centered at $v_{\rm
LSR}$~$\simeq$~10 km~s$^{-1}$, and a broad component ($\Delta v \simeq$ 20
km~s$^{-1}$), centered at $v_{\rm LSR} \simeq 23$ km~s$^{-1}$ 
(see Fig. \ref{fig_co_no_residual}). Our
observations with the multibeam HERA array reveal that, while the emission
of the narrow component is present over a large region ($\gg1'$), the
broad component is only detected toward the position of K~3-35. Given its
line width and angular extension, we conclude that the narrow component is
most probably of interstellar origin, while the broad component is most likely 
associated with the PN K~3-35. In order to better isolate the CO emission 
associated with K~3-35, a Gaussian fit to the narrow component was subtracted 
from the overall spectrum. 

In Figure \ref{fig_co_iram_nobeyama} we show the resulting CO
($J=2\rightarrow1$) spectrum toward K~3-35, after subtraction of the
narrow component. In Table~\ref{table_1} we give the line parameters 
of the broad component associated with K~3-35. 
The values of the central velocity and line width are similar to those
reported previously by \cite{DB96} using the Kitt Peak 12 m telescope and
more recently by \cite{Hugg05} using the Pico Veleta 30 m telescope. Our
line intensity is higher than the value reported by \cite{Hugg05}, and 
this can be due to slight differences in the pointing, given the small
size of the 30 m beam at 1.3 mm. Figure \ref{co_mosaic} shows a CO
($J=2\rightarrow1$) spectra mosaic (after subtraction of the narrow
component at 10 km s$^{-1}$) of a region of $1'\times1'$, centered on
K~3-35. This mosaic shows that broad emission is coming from within a region
smaller than $\sim 20''$, providing a constraint on the angular size of the
K~3-35 molecular envelope.

We also have detected weak broad CO ($J=1\rightarrow0$) emission toward K~3-35
using the Nobeyama 45 m telescope (see lower panel of 
Fig.\ \ref{fig_co_iram_nobeyama} and Table~\ref{table_1}). 
An interstellar narrow CO ($J=1\rightarrow0$) component was 
also present and subtracted as in the CO ($J=2\rightarrow1$) data. 
The central velocity of the broad emission is $v_{\rm LSR}$~$\simeq$~27~km~s$^{-1}$ 
and the line width is $\Delta v\simeq$ 10~km~s$^{-1}$. 
This is the first time that the CO ($J=1\rightarrow0$) 
transition has been reported in association with this PN.
Although the line is only marginally detected (3-$\sigma$) and more sensitive
observations are necessary to confirm the detection, it is located at the 
right velocity, suggesting a true association with K~3-35. We attribute the 
apparent difference in line widths between the CO ($J=2\rightarrow1$) and CO 
($J=1\rightarrow0$) lines to the modest signal-to-noise ratio of the CO 
($J=1\rightarrow0$) detection.

The CO emission is very useful for investigating the physical parameters of
the molecular gas associated with the PN K~3-35.  An estimate of the
kinetic temperature of the gas can be obtained from the ratio of the
CO ($J=2\rightarrow1$) to CO ($J=1\rightarrow0$) transitions. Assuming that the
level populations are well thermalized, that the emission is optically
thin, and neglecting the correction for departures from the Rayleigh-Jeans
regime, the kinetic temperature can be approximated by the following
equation,
\begin{equation}
T_{k}=-\frac{11.06}{\ln(R/4)},
\end{equation} 
 where $R$ is the ratio of the CO ($J=2\rightarrow1$) to CO
($J=1\rightarrow0$) velocity integrated line intensities corrected by the
difference in beam sizes. If the emission is unresolved in both
transitions, $R=(\theta_{2\rightarrow1}/\theta_{1\rightarrow0})^2 \int
T_{\rm MB}({\rm CO; 2\rightarrow1})\,dv/\int T_{\rm MB}({\rm CO;
1\rightarrow0})\,dv$, being $\theta$ the FWHM of the telescope main beam.
Using the integrated intensities of our 30~m and 45~m CO observations, we
derived a value of $20\pm6$ K for the kinetic temperature of the molecular
gas in K~3-35. Analysis of previous observations led to a higher kinetic 
temperature estimate ($T_{k}\geq 120$ K; Dayal \& Bieging 1996). However,
from our observations, and taking properly into account the difference in the
telescope beam sizes, we derive for K~3-35 a lower value for the kinetic
temperature, which is more similar to other kinetic temperature
determinations in young PNe, ranging typically from 25 to 60 K 
(Bachiller et al. 1997).

The CO column density and molecular mass can be also derived from the CO
observations assuming LTE conditions and that the emission is optically
thin. For this, we use the CO ($J=2\rightarrow1$) line since its spectrum 
is of better quality (see Fig. \ref{fig_co_iram_nobeyama}). The beam 
averaged CO column density can be obtained using the following relation,
\begin{equation}
\left[\frac{N(\mbox{CO})}{\mbox{cm}^{-2}}\right]
= \frac{1.1\times10^{13}\,\,T_k}{(e^{-5.54\,{\rm
K}/T_k}-e^{-16.60\,{\rm K}/T_k}) [(e^{11.06\,{\rm K}/T_k}-1)^{-1}-0.02]}
\left[\frac{\int{T_{\rm MB}}({\rm CO; 2\rightarrow1})\;\mbox{d}v}
{\mbox{K km s}^{-1}}\right].
\end{equation}

For a kinetic temperature $T_k$ = 20 K, we obtain a beam averaged CO
column density $N$(CO) = 5.1 $\times$ 10$^{15}$ cm$^{-2}$ for the K~3-35
envelope. The result is weakly dependent on the precise value of $T_k$
adopted (for $T_{k}$ $\simeq$ 100 K, $N$(CO) would be only a factor of
2.6 higher).

The molecular mass of the envelope can be estimated using the following
expression,
 \begin{equation}
\left[\frac{M_m}{M_{\odot}}\right]=2.9\times10^{-25}\left[\frac{\theta}{\rm
arcsec}\right]^{2}\left[\frac{D}{\rm kpc}\right]^{2}\left[\frac{N(\rm
CO)}{\rm cm^{-2}}\right]X_{\rm CO}^{-1},
 \end{equation}
 where $\theta$ is the FWHM of the telescope beam, $D$ is the distance to
the source, and $X_{\rm CO}$ is the abundance of CO relative to H$_2$.
Using a representative value for PNe, $X_{\rm CO}\simeq 3\times10^{-4}$
(Huggins et al. 1996), and adopting a distance of $D=5$ kpc to K~3-35, we
estimate a molecular mass for the envelope of $M_m$ $\simeq$ 0.017
$M_{\odot}$. This result can be easily scaled for different values of
$T_k$, $D$, and $X_{\rm CO}$ using the expressions given above.

\subsection{HCO$^{+}$}

HCO$^+$ ($J=1\rightarrow0$) emission was detected toward K~3-35 with the
20 m Onsala telescope. This is the first time that HCO$^+$ emission has
been reported in association with this PN. The observed spectrum is shown
in Figure~\ref{fig_hcom} and the line parameters are given in
Table~\ref{table_1}. The mean velocity of the HCO$^+$ emission is $v_{\rm
LSR}\simeq28$~km~s$^{-1}$ and the line width is $\Delta v \simeq
20$~km~s$^{-1}$. These values are similar to those obtained from the CO
$(J=2\rightarrow1)$ observations reported by \cite{DB96}, \cite{Hugg05},
and in the present work, supporting that the detected HCO$^{+}$ emission
is associated with the planetary nebula K~3-35.

Assuming LTE conditions and that the emission is optically thin, we can estimate 
the beam averaged HCO$^{+}$ column density from our observations using the
following equation,
 \begin{equation}
 \left[\frac{N({\rm HCO^+})}{\mbox{cm}^{-2}}\right]
 = \frac{7.9\times10^{10}\;\;T_{k}}{(1-e^{-4.29\,{\rm K}/T_k})
[(e^{4.29.\,{\rm K}/T_k}-1)^{-1}-0.26]}
\left[\frac{\int{T_{\rm MB}({\rm HCO^+; 1\rightarrow0})\;\mbox{d}v}}
{\mbox{K km s}^{-1}}\right].
 \end{equation}
 
For a kinetic temperature $T_k=20$~K, the resulting 
beam averaged HCO$^{+}$ column density is $N$(HCO$^{+}$) $\simeq$ 6.9 $\times$
10$^{11}$ cm$^{-2}$. This value implies for K~3-35 an HCO$^+$ abundance
relative to CO, $\rm [HCO^+/CO]=1.9\times10^{-3}$, where we have corrected 
for the difference in the beam sizes, and an HCO$^+$
abundance relative to H$_2$, $X_{\rm HCO^+}=5.7\times10^{-7}$, assuming
that $X_{\rm CO}=3\times10^{-4}$.

\section{Discussion}

The presence of water maser emission led to the suggestion that K~3-35 is
an extremely young PN (Miranda et al. 2001), given that water molecules
are expected to survive only for a short time ($\la 100$ yr) during the PN
phase. However, the existence of these molecules in the envelope may not
be an unambiguous indicator of youth, since dense molecular clumps could
be protecting the water molecules from photo-dissociation and let them
survive for a longer time. An alternative way to probe the evolutionary
status of PNe is to analyze their molecular content. It is known that the
molecular abundances in protoplanetary nebulae (PPNe) envelopes change as
the core begins to produce UV radiation and the star enters in the PN
phase. In particular, it has been observed that the HCO$^{+}$ abundance
increases rapidly when the star goes from the PPN to PN phase (Bachiller
et al. 1997; Josselin \& Bachiller 2003), providing some indications on
the evolutionary state. Furthermore, Huggins et al. (1996) found that the
ratio of molecular to ionized mass, $M_m/M_i$, is inversely correlated
with the radius of the nebula, and therefore it can be taken as an
indicator of the evolutionary stage of the PN, almost independent of the
distance. 

In order to probe the evolutionary status of K~3-35 and compare it with 
other objects, we estimated the ratio $M_m/M_i$ for K 3-35. The molecular 
mass for K 3-35 was obtained from our results in \S 3.2. To estimate the 
mass of ionized gas in K~3-35, we have followed the formulation of Mezger 
\& Henderson (1967) and used the 3.6 cm radio continuum data of Miranda et 
al. (2001). The value for the ionized mass is $M_{i}$ $\simeq 9 
\times10^{-3}~M_\odot$, resulting in a ratio of molecular to ionized mass, 
$M_m/M_i\simeq1.9$ for K~3-35. This value for the ratio of molecular to 
ionized mass in K~3-35 is comparable to the values found for other young 
PNe (e.g. IC 5117) with kinematical ages of a few hundred years (Miranda 
et al. 1995). In order to further test the evolutionary status of K~3-35, 
we plot in Figure 5 the HCO$^{+}$ abundance as a function of the ratio of 
molecular to ionized mass for K 3-35, as well as for a sample of PPNe and 
PNe taken from Bachiller et al. (1997) and Josselin \& Bachiller (2003). 
Figure 5 shows that the HCO$^{+}$ abundance in K~3-35 is $\sim$ 500 times 
higher than the abundance in the PPN CRL 2688, and that it has reached a 
value comparable (in fact, somewhat higher) to those found in more evolved 
PNe.

The HCO$^{+}$ traces molecular gas with densities of order of 
$\sim$10$^{5}$~cm$^{-3}$, such as that forming dense molecular clumps 
embedded in the ionized material (Huggins et al. 1992). It is feasible 
that these clumps could be protecting other molecules, such as H$_2$O, 
from the ionizing radiation. Therefore, the detection of HCO$^{+}$ in 
K~3-35 suggests that dense molecular clumps could be responsible for the 
shielding of the water molecules present in this nebula. Higher angular 
resolution observations at millimeter and submillimeter wavelengths would 
be very valuable to confirm the clumpiness in K~3-35.

In summary, our study of K~3-35 indicates that the molecular data on this 
PN are consistent with this source being a relatively young (a few hundred 
years) PN, where the survival of water molecules may be favored by the 
presence of dense molecular clumps that could be protecting the water 
molecules from photo-dissociation.
 

\section{Conclusions}

We have carried out a survey for molecular emission toward the young 
planetary nebula K~3-35. As a result of this survey, we detected for the 
first time HCO$^{+}$ ($J=1\rightarrow0$) and CO ($J=1\rightarrow0$) 
emission toward this PN. The emission appears in a velocity range that is 
in agreement with that of the broad CO ($J=2\rightarrow1$) emission 
reported in previous studies, as well as in the present work. We have also 
mapped the CO ($J=2\rightarrow1$) line, showing that the emission is 
compact and centered on K~3-35, confirming the association with the PN, 
and setting an upper limit for the angular size of the molecular envelope 
around K~3-35 of $\la$~$20''$.  We have used the ratio of the CO 
($J=2\rightarrow1$) and CO ($J=1\rightarrow0$) lines to obtain a better 
constrained estimate of the kinetic temperature in the K~3-35 molecular 
envelope, which turns out to be $\sim$ 20$\pm$6 K. From our observations 
we have found that the HCO$^{+}$ abundance in K~3-35 is $\sim 6 \times 
10^{-7}$, which is similar to values found in other young PNe. This 
result, along with the value for the ratio of molecular to ionized mass, 
$M_m/M_i$ $\simeq$ 1.9, suggest that K~3-35 might not be an extremely 
young ($<100$ yr) PN but a somewhat more evolved one (a few hundred years) 
where the presence of water maser emission could be favored by a shielding 
mechanism that prevents the water molecules from being dissociated. Since 
it is believed that the HCO$^{+}$ emission could be arising in dense 
clumps of molecular material embedded in the ionized gas, its detection in 
K~3-35, a source of water maser emission, provides an important clue to 
understand the shielding mechanism that could be protecting the water 
molecules from the radiation of the central star. Interferometric 
molecular observations at millimeter and submillimeter wavelenghts would 
be very valuable to determine the morphology of the molecular envelope and 
to investigate the clumpiness of the HCO$^+$ emission in this remarkable 
young planetary nebula.

\acknowledgments

RF-H, YG, LL, and DT acknowledge the support from DGAPA, UNAM and CONACyT, 
Mexico. GA, MO, and JMT acknowledge partial financial support from grant 
AYA2005-08523-C03 of the Spanish MEC (co-founded with FEDER funds). LFM is 
supported partially by grant AYA 2005-01495 of the Spanish MEC (co-founded 
with FEDER funds). GA, MO and LFM acknowledge support from Junta de 
Andaluc\'\i a. JMT acknowledges the warm hospitality of the UK Astronomy 
Technology Centre, Royal Observatory Edinburgh, during his sabbatical 
stay.

\clearpage

\begin{figure}
\begin{center}
\includegraphics[angle=0,scale=.6]{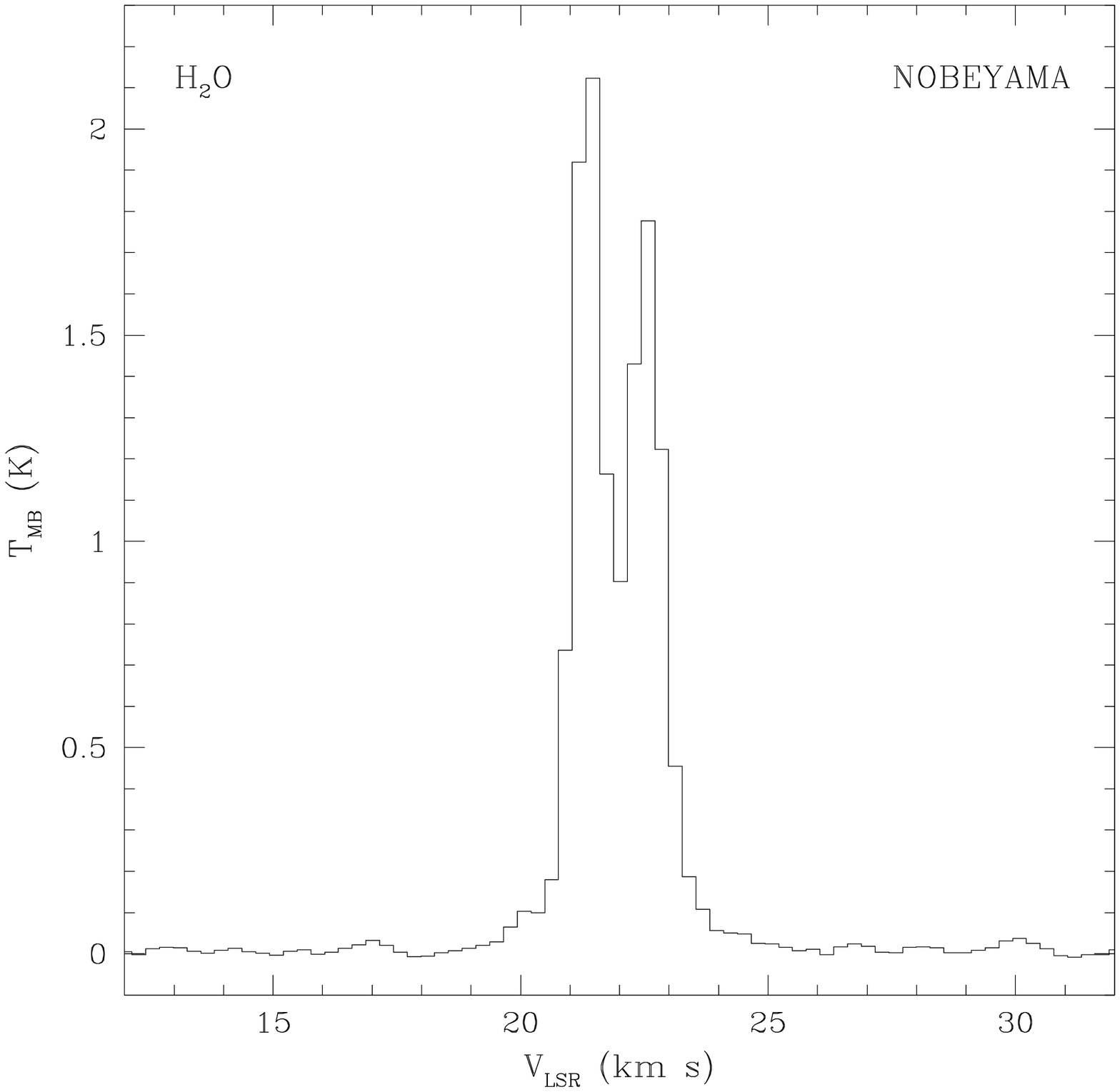}
 \caption{Water maser spectrum toward K~3-35 observed on March 9, 2003 
with the Nobeyama 45 m telescope. The spectrum has been smoothed to a 
resolution of 0.25 km~s$^{-1}$ and a 1st order polynomial baseline has 
been subtracted. The rms noise in the off-line channels is 0.01 K. Two 
components at LSR velocities of 21.4 and 22.5 km s$^{-1}$ can be 
identified. \label{fig3}}
 \end{center}
\end{figure}

\clearpage

\begin{figure}
\begin{center}
\includegraphics[angle=0,scale=.6]{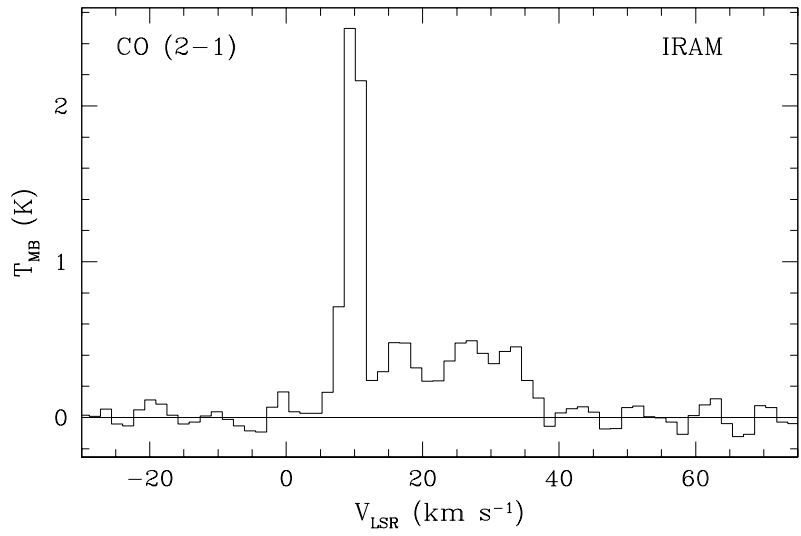}
 \caption{ CO ($J=2\rightarrow1$) spectrum toward K~3-35
observed with the IRAM 30 m telescope. The spectrum has been smoothed to a
resolution of 1.6 km~s$^{-1}$ and a 2nd order polynomial baseline has been
subtracted. The emission has two velocity components,
a narrow component ($\Delta v~\simeq$~5~km~s$^{-1}$), centered at $v_{\rm
LSR}$~$\simeq$~10 km~s$^{-1}$, and a broad component ($\Delta v \simeq$ 20
km~s$^{-1}$), centered at $v_{\rm LSR} \simeq 23$ km~s$^{-1}$.
\label{fig_co_no_residual}}
 \end{center}
\end{figure}

\clearpage

\begin{figure}
\begin{center}
\includegraphics[angle=0,scale=0.65]{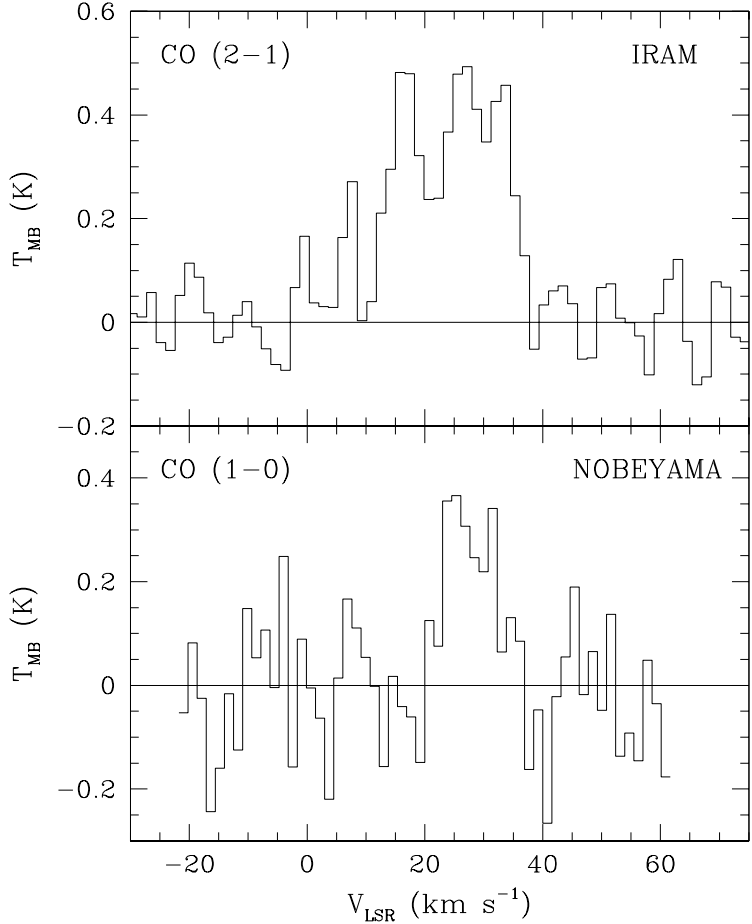}
 \caption{Upper panel: CO ($J=2\rightarrow1$) spectrum toward K~3-35
observed with the IRAM 30 m telescope. The spectrum has been smoothed to a
resolution of 1.6 km~s$^{-1}$ and a 2nd order polynomial baseline has been
subtracted. The narrow, interstellar component at $v_{\rm LSR}$~$\simeq$~10
km~s$^{-1}$ (Fig. 2) has also been subtracted (see \S 3.2). The rms noise in the off-line
channels is $\sim$ 0.063 K. The line emission is centered at $v_{\rm LSR}$
$\simeq$~23~km~s$^{-1}$, with a line-width $\Delta v$ $\simeq$ 20 km~s$^{-1}$. 
Lower panel: CO ($J=1\rightarrow0$) spectrum toward K~3-35
observed with the Nobeyama 45 m telescope. The spectrum has been smoothed
to a resolution of 1.5 km~s$^{-1}$ and a 3rd order polynomial baseline has been
subtracted. The narrow, interstellar component at $v_{\rm LSR}$~$\simeq$~10
km~s$^{-1}$ has also been subtracted. The rms noise in the off-line channels is 
$\sim$ 0.1 K. The line emission is centered at $v_{\rm LSR}$~$\simeq$~27~km~s$^{-1}$, 
with a line-width $\Delta v$ $\simeq$ 10 km~s$^{-1}$.
\label{fig_co_iram_nobeyama}}
 \end{center}
\end{figure}

\clearpage

\begin{figure}
\begin{center}
\includegraphics[angle=0,scale=0.9]{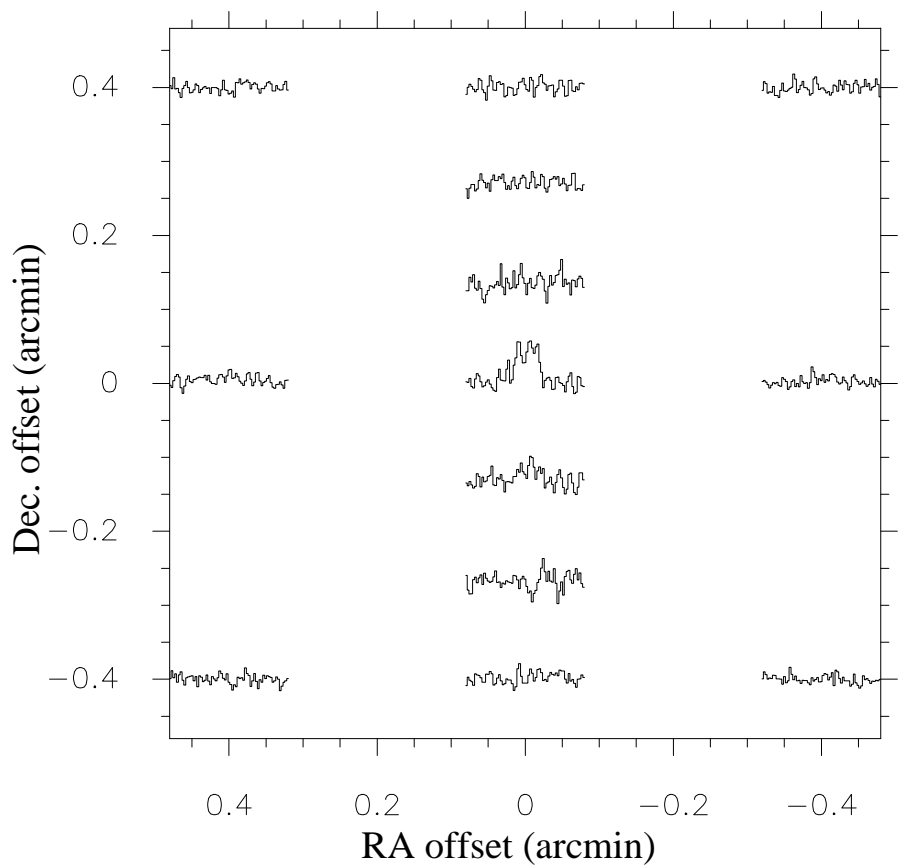}
 \caption{Mosaic of CO ($J=2\rightarrow1$) spectra in the region around
K~3-35 obtained with the IRAM 30 m telescope.  A narrow, interstellar
component at $v_{\rm LSR}$~$\simeq$~10 km~s$^{-1}$, present in all the
positions, has been subtracted from the spectra. The (0,0) offset
position corresponds to the position of K~3-35, RA(J2000)=$19^h 27^m 44^s$, 
Dec(J2000)=$21^\circ 30' 03''$. Note that the emission arises from a 
compact region toward K~3-35. \label{co_mosaic}}
 \end{center}
\end{figure}

\clearpage

\begin{figure}
\begin{center}
\includegraphics[angle=0,scale=0.6]{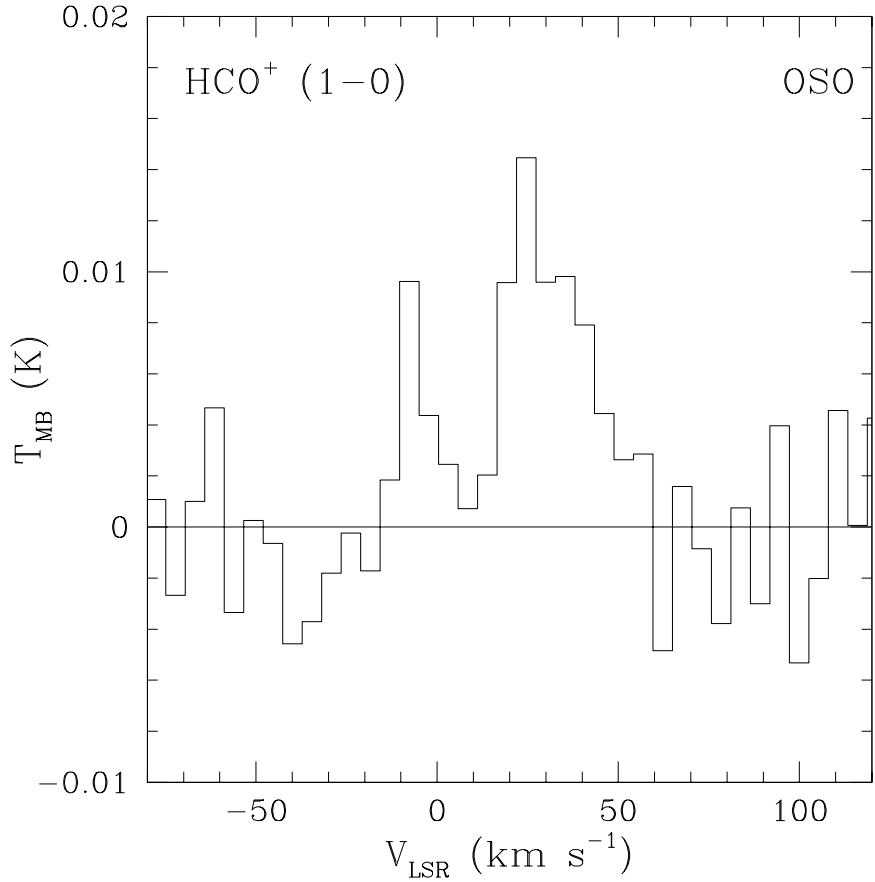}
\caption{HCO$^{+}$ ($J=1\rightarrow0$) spectrum toward K~3-35 observed
with the Onsala 20~m telescope. The spectrum has been smoothed to a
resolution of 5.4 km~s$^{-1}$ and a 2nd order polynomial baseline has been
subtracted. The rms noise in the off-line channels is $\sim$ 0.003 K.  
The line emission is centered at $v_{\rm LSR}$ $\simeq$~28~km~s$^{-1}$,
with a line-width $\Delta v$ $\simeq$ 20 km~s$^{-1}$.\label{fig_hcom}}
\end{center}
\end{figure}

\clearpage

\begin{figure}
\begin{center}
\includegraphics[angle=0,scale=0.6]{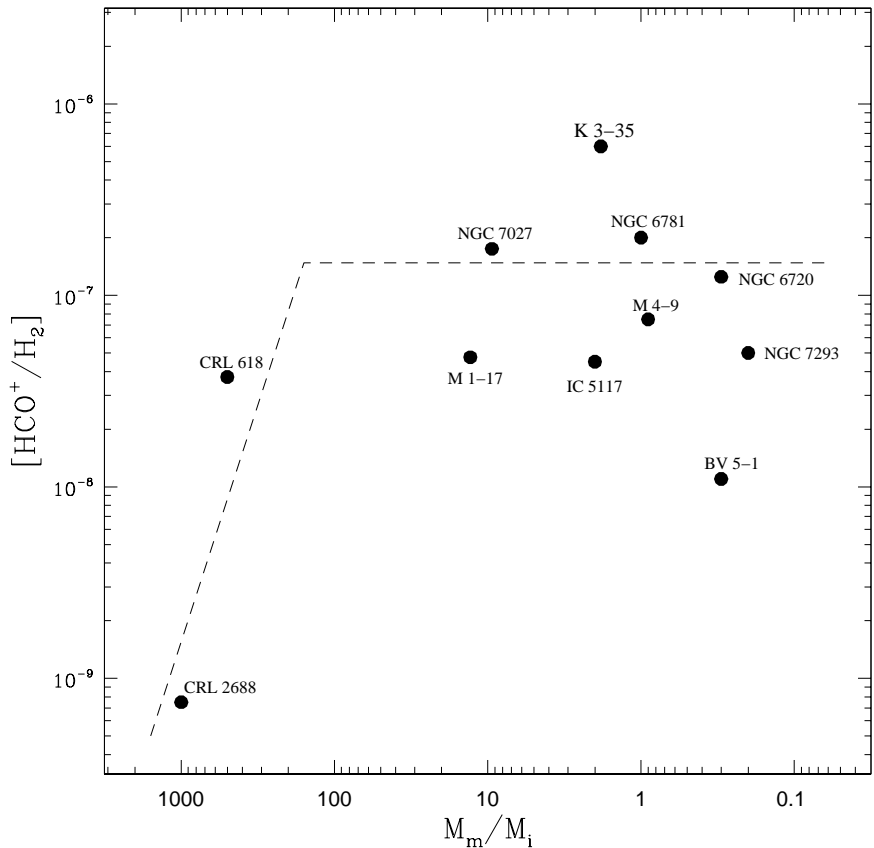}
 \caption{HCO$^{+}$ abundance as a function of the ratio between molecular
and ionized mass, for several PPNe and PNe. The $M_m/M_i$ ratio has been
found to decrease with the age of the PN (Huggins et al. 1996).  
Therefore, the plot illustrates the time evolution of the HCO$^+$
abundance. The dashed line represents the average HCO$^+$ abundance for
PPNe and PNe. Note the significant increase in the HCO$^+$
abundance from the PPN to the PN phase. Data points are taken from
Bachiller et al. (1997) and Bachiller \& Josselin (2003), except for
K~3-35 (this work).  
\label{age_abundance}}
\end{center}
\end{figure}

\clearpage
\begin{deluxetable}{cclllccccccc}
\rotate
\tabletypesize{\tiny}
\tablecolumns{12} 
\tablecaption{Single Dish Observations toward K~3-35 \label{table_1}}
\tablewidth{0pc}
\tablehead{
\colhead{Telescope} & \colhead{Date} & \colhead{Molecule} & 
\colhead{Transition} & \colhead{Frequency} 
&\colhead{$\eta_{\rm MB}$\tablenotemark{a}} 
&\colhead{$\Gamma$\tablenotemark{b}} &\colhead{$\Delta$ch\tablenotemark{c}} & \colhead{rms ($T_{\rm MB}$)\tablenotemark{d}} &\colhead{$\int{T_{\rm MB}}$d$v$\tablenotemark{e}}& \colhead{$v_{\rm LSR}$\tablenotemark{f}} &\colhead{$\Delta v$\tablenotemark{g}}\\
 \colhead{} & \colhead{} & \colhead{} & \colhead{} &\colhead{(GHz)} &\colhead{} &\colhead{(K Jy$^{-1}$)}& \colhead{(km s$^{-1}$)} & \colhead{(K)} &\colhead{(K km s$^{-1}$)} & \colhead{(km s$^{-1}$)}&\colhead{(km s$^{-1}$)} 
}
\startdata
20m OSO & Feb-Mar 2003 & SiO &$v = 1$, $J = 2 \rightarrow1$& \phantom{0}86.2434420 &0.65 &0.045 &0.174 & 0.024 &\dots & \dots& \dots\\
''& '' & H$^{13}$CO$^{+}$&$J= 1\rightarrow0$& \phantom{0}86.7542940 &0.64 &0.045 &0.173& 0.024 &\dots &\dots &\dots\\
''& '' & HCN &$J = 1\rightarrow0$, $F = 2\rightarrow1$& \phantom{0}88.6318473 &0.63 &0.043 &0.169 & 0.017 &\dots &\dots &\dots\\
''& '' & HCO$^{+}$&$J = 1\rightarrow0$& \phantom{0}89.1885180 &0.63 &0.043 &0.168 & 0.012 &0.33$\pm$0.03& 28 & 20\\
''& '' & HNC &$J = 1\rightarrow0$, $F = 2\rightarrow1$& \phantom{0}90.6635430 &0.61 &0.043 &0.165 &0.024 & \dots& \dots &\dots\\
''& '' & CH$_3$OH&$J = 8_{0}\rightarrow7_{1} A^{+}$ & \phantom{0}95.1694400 &0.58 &0.040 &0.158 &0.028 &\dots & \dots &\dots\\
''& '' & HC$_5$N &$J = 36\rightarrow35$& \phantom{0}95.8503370 &0.58 &0.040 &0.156 & 0.030& \dots& \dots &\dots\\
''& '' & CS &$J = 2\rightarrow1$& \phantom{0}97.9809680 &0.56 &0.038 
&0.153 &0.027 &\dots & \dots &\dots\\
''& '' & HC$_3$N &$J = 12\rightarrow11$& 109.173634 &0.47 &0.034 &0.137 & 0.039 & \dots& \dots &\dots\\
''& '' & $^{13}$CO &$J = 1\rightarrow0$& 110.201353 &0.47 &0.033 &0.068 & 
0.045 & \dots & \dots &\dots\\
''& '' & CN &$J = 1\rightarrow0$& 113.490982 &0.44 &0.032 &0.132 & 0.044 & \dots& \dots &\dots\\
45m Nobeyama & Mar 09, 2003 & H$_{2}$O &$J = 6_{16}\rightarrow5_{23}$& 
\phantom{0}22.235080 &0.81 &0.357 &0.250& 0.010 & 1.5 $\pm$0.1 & 21.4 &0.72\\
& & & & & & &0.250 & 0.010 &1.6$\pm$0.1 & 22.5&0.88\\
''&  Mar 2003& NH$_{3}$ &$(J,K) = (1,1)$&\phantom{0}23.695110 &0.81 &0.357 & 0.250 & 0.010 &\dots& \dots&\dots\\
''& ''& CO &$J = 1\rightarrow0$& 115.271204 &0.46 &0.196 &0.083 &0.200 
&2.7$\pm$0.4 & 27& 10\\
30m IRAM & Apr 01, 2003 & CO & $J = 2\rightarrow1$ & 230.537990 &0.58 &0.105 & 0.406 &0.088 & 10.2$\pm$0.4 & 23 & 20\\
\enddata

\tablenotetext{a}{Main beam efficiency. The main beam brightness temperature is obtained as $T_{\rm MB}=T'_A/\eta_{\rm MB}$, where $T'_A$ is the antenna temperature corrected for the atmospheric attenuation.}
\tablenotetext{b}{Sensitivity of the telescope. The flux density can be 
obtained as $S_\nu=\eta_{\rm MB}T_{\rm MB}/\Gamma$.}
\tablenotetext{c}{Channel width.}
\tablenotetext{d}{1-$\sigma$ rms noise per channel (in $T_{\rm MB}$ scale).}
\tablenotetext{e}{Zero-order moment (velocity integrated intensity) of the line emission.}
\tablenotetext{f}{First-order moment (intensity weighted mean $v_{\rm LSR}$) of the line emission.}
\tablenotetext{g}{Equivalent line width, obtained from the second-order moment. For a Gaussian line 
profile, this value corresponds to the FWHM.}

\end{deluxetable}

\end{document}